\documentclass[sigconf]{acmart}

\AtBeginDocument{%
  }

\copyrightyear{2025}
\acmYear{2025}
\setcopyright{acmlicensed}\acmConference[MECC '25]{2nd International Workshop on MetaOS for the Cloud-Edge-IoT Continuum }{March 30-April 3 2025}{Rotterdam, Netherlands}
\acmBooktitle{2nd International Workshop on MetaOS for the Cloud-Edge-IoT Continuum (MECC '25), March 30-April 3 2025, Rotterdam, Netherlands}
\acmDOI{10.1145/3721889.3721926}
\acmISBN{979-8-4007-1560-0/2025/03}
\begin{document}

\title{Identity and Access Management for the Computing Continuum}

\author{
Chalima~Dimitra~Nassar~Kyriakidou 
Athanasia~Maria~Papathanasiou 
Vasilios~A.~Siris}
\email{{dnassar,papathanasiou,siris}@excid.io}
\affiliation{
  \institution{Athens~Univ.~of~Economics~\&~Business}
  \city{Athens}
\country{Greece}
}
\affiliation{%
\institution{ExcID~P.C.}
\city{Athens}
\country{Greece}
}

\author{
Nikos~Fotiou George~C.~Polyzos}
\email{{fotiou,polyzos}@excid.io}
\affiliation{%
  \institution{ExcID~P.C.}
  \city{Athens}
  \country{Greece}
}

\author{Eduardo Cánovas Martínez
Antonio Skarmeta}
\email{{eduardo.canovas1,skarmeta}@um.es}
\affiliation{%
  \institution{Universidad de Murcia}
  \city{Murcia}
  \country{Spain}
}

\renewcommand{\shortauthors}{Kyriakidou et al.}

\begin{abstract}
The computing continuum introduces new challenges for access control due to its dynamic, distributed, and heterogeneous nature. In this paper, we propose a Zero-Trust~(ZT) access control solution that leverages decentralized identification and authentication mechanisms based on Decentralized Identifiers~(DIDs) and Verifiable Credentials~(VCs). Additionally, we employ Relationship-Based Access Control~(ReBAC) to define policies that capture the evolving trust relationships inherent in the continuum. Through a proof-of-concept implementation, we demonstrate the feasibility and efficiency of our solution, highlighting its potential to enhance security and trust in decentralized environments.  
\end{abstract}




\maketitle

\section{Introduction}
The evolution of the computing continuum, encompassing cloud, edge, and decentralized infrastructures, has introduced new challenges in security, privacy, and trust management. Modern computing ecosystems are increasingly composed of heterogeneous devices, ranging from powerful cloud servers to resource-constrained edge nodes, all of which require seamless and secure access management~\cite{zhu2014rbac}. 
Traditional \emph{Identity and Access Management}~(IAM) models, which rely on centralized identity providers and predefined access control policies, struggle to meet the dynamic requirements of the cloud-to-edge continuum~\cite{indu2018identity, mohammed2018identity, mohammed2019cloud, alsirhani2022advanced}. 

A key challenge in cloud-to-edge security is the ability to enforce real-time access control that adapts to contextual changes, such as dynamic user roles, network conditions, and device mobility. 
Existing access control models often assume static environments, making them insufficient for managing continuously evolving relationships between users, devices, and services. 
Furthermore, access control enforcement must be positioned closer to protected resources—at the edge—where data and devices are physically exposed and more vulnerable to security threats. 

Moreover, computing ecosystems with diverse stakeholders require flexible trust models that enable secure collaboration between independent domains, while maintaining local policy autonomy. 
Trust relationships must be dynamically established and enforced, ensuring sovereignty over local access control decisions. 
This need is particularly challenging in environments where access policies must be validated at enforcement time~\cite{kyriakidou2023decentralized}. Finally, preventing unauthorized delegation of permissions, such as access token sharing, is crucial to maintaining security, particularly in loosely connected environments where centralized enforcement is infeasible.

Conventional access control models, such as \emph{Attribute-based Access Control}~(ABAC)~\cite{gupta2020attribute, ruj2014attribute, xue2019attribute, sookhak2017attribute} and \emph{Role-based Access Control}~(RBAC)~\cite{zhou2013achieving, butt2023optimized, zhou2015trust, xu2021role}, which rely on static attributes and roles, fail to capture the evolving relationships between entities such as IoT devices, users, and computing nodes, and thus lack the agility required for enforcing security policies in cloud-to-edge scenarios.
To address these challenges, we propose an IAM solution that enables decentralized identity management and continuous authentication and authorization across distributed computing environments. Our approach leverages the \emph{Zero-Trust}~(ZT) security model, utilizing \emph{Decentralized Identifiers}~(DIDs), \emph{Verifiable Credentials|}~(VCs), and contextualized authorization to enhance security, while preserving system efficiency. Specifically, we propose a decentralized IAM framework, making the following contributions:
\begin{itemize}
    \item Decentralized identity management and trust management: We leverage DIDs and VCs for user-centric and cryptographically verifiable identities to support flexible trust models in the computing continuum. Our trust models can capture the multitude of diverse stakeholders with different interests and relations that mat exist in continuum.  
    \item Adaptive and contextual access control: We leverage the \emph{Relationship-based Access Control}~(ReBAC) model, allowing access control policies to dynamically adapt based on real-time relationships between entities. 
    \item Continuous and secure authentication and authorization: We ensure long-lasting yet continuously re-evaluated authorization decisions, preventing unauthorized delegation. In this way, our solution mitigates the risk of unauthorized access token sharing, ensuring robust security even in loosely connected environments.
\end{itemize}

The remainder of the paper is organized as follows. In Section 2, we provide the necessary background that will be used for designing and implementing our IAM solution, along with a comparison with related systems that have been proposed in the literature. In Section 3, we provide the design of our solution, referring to the entities, the trust model and the architecture of our system, as well as describing a simple use case scenario. In Section 4, we provide a detailed explanation of our implementation, while in Section 5 we evaluate our solution in terms of performance. Finally, Section 6 concludes our work.

\section{Background and Related Work}

\subsection{Decentralized Identifiers}
Decentralized Identifiers (DIDs)~\cite{W3C-DID-Core} are a W3C recommendation designed to enable self-sovereign and verifiable digital identities without relying on centralized authorities. A DID is a globally unique identifier in the form of a URL, which can be resolved to a DID \emph{document}. This DID document contains essential information, including public keys, service endpoints, and authentication mechanisms, allowing entities to prove ownership and establish trust without intermediaries. Unlike traditional identifiers managed by centralized entities, DIDs are designed to be user-controlled, enhancing privacy, security, and interoperability across different systems and platforms.

The DID specification allows for multiple DID \emph{methods} to coexist, providing flexibility in how DIDs are created, managed, and resolved. The primary distinction between these methods lies in where the DID document is stored and how its resolution process is implemented. Some methods rely on blockchain or distributed ledger technologies (DLTs) to ensure tamper resistance and decentralization, while others use peer-to-peer networks or traditional databases. This diversity in DID methods allows for a broad range of use cases, from decentralized authentication to secure communication and verifiable credentials, making DIDs a fundamental building block for decentralized identity ecosystems.

\subsection{Verifiable Credentials}
A Verifiable Credential (VC) is defined~\cite{Sporny2024} as a digital document that represents the same claims with a physical document, but in a more tamper-proof and trustworthy way compared to its physical counterpart. A VC subject is defined as the entity upon which claims are made, and therefore a subject may for instance refer to a person, an organization or an IoT device. The holder of the VC is typically equivalent to the VC subject, but there are cases in which they refer to different entities, such as when a parent holds the VCs of her child. The W3C VC Ecosystem 
consists of an issuer, a holder and a verifier. Initially, the issuer is responsible for creating and signing a VC, which corresponds to the holder's claims. The VC includes a unique identifier that is used to correlate the holder with the credential, such as a DID or a public key. 

In case holders wish to present a VC to a verifier, they can combine one ore more credentials to a \emph{Verifiable Presentation}~(VP). Therefore, the VP includes certain VCs signed by the issuer, along with the holder's signature on the presentation. Upon receiving the VP, a verifier should validate both the issuer's and holder's signatures using their corresponding public keys. 


\subsection{Relationship-based Access Control}
Relationship-based Access Control (ReBAC) is an access control model in which permissions are determined based on relationships between entities. Our solution is based on Google’s Zanzibar ReBAC system~\cite{pang2019zanzibar}. 
ReBAC allows the specification of an \emph{authorization model} using \emph{types} and the possible \emph{relations} among these types. Entities such as users, devices, services, or resources can be categorized into types, and relationships between them define access permissions. Additionally, ReBAC enables the definition of explicit relationships, where certain permissions imply others. For example, if a user has a "write" relationship with a resource type, they are also granted a "read" relationship by default. This structured approach enhances flexibility and ensures that access control policies reflect real-world trust relationships dynamically.

Beyond explicit relationships, ReBAC also supports relation inheritance, where permissions granted at a higher level automatically propagate to related entities. For instance, if a user has a "write" relationship with a type representing a group of resources (e.g., a "smart home"), they also inherit the "write" relationship for individual resources within that group (e.g., smart lights or security cameras). Access control rules are defined by administrators in the form of tuples [subject, relation, object], specifying which entity holds a given relationship with another entity. When an access request is made, the system evaluates queries of the form "Does subject U have relation R with object O?" to determine whether access should be granted.

\subsection{Related Work}
Various access control models have been proposed to address the diverse requirements of cloud and edge environments. In particular,
access control models, such as \emph{Mandatory Access Control}~(MAC)~\cite{meghanathan2013review, taubmann2016cloudphylactor}, ABAC~\cite{gupta2020attribute, ruj2014attribute, xue2019attribute, sookhak2017attribute, 10602646} and RBAC~\cite{zhou2013achieving, butt2023optimized, zhou2015trust, xu2021role}, have been adapted for cloud environments, but their effectiveness is often limited due to the cloud’s dynamic and multi-tenant nature. In an attempt to overcome these challenges, the authors in~\cite{mehraj2021flexible} introduce a dynamic authorization system that integrates roles, tasks, and user trust levels to offer scalable and fine-grained access control in cloud settings. Another related method proposed by Habiba~\textit{et al.}~\cite{habiba2016new} involves a dynamic access control framework that incorporates policy conflict resolution and authorization validation to enhance security in cloud computing.

Moreover, edge and cloud computing presents additional challenges due to its decentralized structure and the necessity for real-time data processing. As a result, access control mechanisms in these environments must be both adaptable and efficient to accommodate diverse user behaviors. In this context, the authors in~\cite{zhang2023dynamic} developed a dynamic access control system that evaluates user behavior trust, combining RBAC with ABAC models to dynamically regulate user permissions. Recent research efforts have also explored hybrid access control models that integrate cloud and edge computing functionalities. One such approach is a bidirectional fine-grained access control scheme designed to facilitate secure data sharing between cloud and edge environments~\cite{cui2021practical}. This model leverages \emph{Attribute-based Encryption}~(ABE) and proxy re-encryption to reinforce security. Additionally, the authors in~\cite{bhatt2021attribute} propose an ABAC model tailored for AWS IoT so as to enable fine-grained access control in cloud-integrated IoT architectures, demonstrating its effectiveness in industrial IoT applications.

Proposed access control mechanisms, including MAC, RBAC and ABAC, encounter limitations in dynamic and distributed environments, such as the cloud-to-edge continuum due to their reliance on predefined roles and attributes, making them inadequate for environments where user roles and trust relationships evolve continuously. To address this challenge, we integrate ReBAC with VCs and DIDs in order to enable context-aware access control that dynamically evaluates permissions based on real-time interactions. Unlike conventional IAM approaches, which depend on static policies and centralized enforcement, our system ensures continuous authentication and fine-grained authorization across heterogeneous systems, aligning with the ZT paradigm, ensuring that no entity is inherently trusted.

\section{Design}
In this section, we present the main entities of our system, the trust relationships between them, a representative use case, as well as the architecture and workflow of our IAM solution for the cloud-to-edge continuum.

\subsection{System Entities and Trust Model}
Our system consists of the following entities:
\begin{itemize}
    \item Asset Owners: Organizations or individuals who own and manage protected resources, such as IoT devices,
    by defining access policies. 
    \item Organizations: Authorized entities, such as universities and enterprises, that issue credentials to their members, such as employees or students, and define their attributes with respect to the organization.
    \item Users: Individuals or automated systems, such as software agents, requesting access to protected resources. Users are authenticated with credentials issued by the organizations they are affiliated with. Thus, their permissions are defined based on the attributes assigned to them by their organizations and the policies set by the owners of the resources they are requesting access to. 
    \item Asset Groups: Logical collection of assets, such as sensors or devices, that can be managed collectively. Assets may consist of one or more protected resources, and operations can be performed on them based on the access control policies defined by asset owners.
\end{itemize}

The aforementioned entities have trust relationships, which are not implicit, but enforced through continuous validation of credentials, attributes, and policies, that define the interactions within the system. 
Using their IdM organizations issue VCs to their users that validate their identities and define their attributes. 

Asset owners, in turn, 
rely on the organization's issued credentials, which must satisfy the access control policies defined for specific resources or asset groups. 
These policies act as an additional layer of protection, ensuring that only users with the appropriate attributes and permissions, validated through the organization, can interact with protected resources. Polices are in the form "Users with attribute A issued by organization O have relation R with asset S".
This means that asset owners retain significant control over their resources and do not fully delegate trust to the organizations.
Organizations do not have unilateral control over access; they can issue credentials, but they cannot override the policies enforced at the access control layer.

Finally, the relationship between asset owners and their resources is one of direct control. 
Asset owners define and enforce the access control policies for their assets and resources, ensuring that interactions with these resources adhere strictly to the owners' policies.
These mechanisms ensure that trust is not blindly placed in any single entity but instead distributed among them, and continuously verified across the interactions of the system.


\subsection{Use Case Scenario}~\label{duc}
To demonstrate the interactions within our system, we consider a smart city deployment, where various IoT devices, such as security cameras, environmental sensors, and energy management systems, collect data that multiple stakeholders may need to access securely. 
This simple scenario involves a Technical University collaborating with the smart city administration to conduct energy consumption research. 
To facilitate this, university researchers require access to real-time energy measurements from city-owned IoT devices. 
However, the city must ensure that access is restricted, verifiable, and revocable, preventing unauthorized entities from retrieving or tampering with sensitive data.

The university registers a researcher in its IdM and issues a VC confirming their affiliation and attributes, such as "cs\_department" and "researcher".
This credential is cryptographically signed and stored in the researcher's digital wallet.
The researcher attempts to access a smart city IoT energy meter via an HTTPS request, which includes a VP generated from their VC, providing the necessary attributes to access the resource.
Her attributes are considered by ReBAC authorization model used for making access control decisions. This model is defined and managed by the smart city administration.
If a researcher leaves the university, the institution updates its IdM to revoke the VC.

From this use case it becomes obvious how are system achieves segregation of trust management: the University is responsibly for managing it user management system, whereas smart city is responsible for managing the authorization model that governs access to the protected resources. 

\subsection{Architecture and Workflow}
The proposed IAM solution leverages ReBAC and VCs to provide decentralized identity and access control across the cloud-to-edge continuum, ensuring that all access requests are explicitly verified before being granted. 
From a high-level perspective, our system consists of an IdM, 
as well as, access control components common to most access control solutions, namely a \emph{Policy Enforcement Point}~(PEP), a \emph{Policy Decision Point}~(PDP), a \emph{Policy Administration Point}~(PAP) and a \emph{Policy Information Point}~(PIP), to ensure dynamic, fine-grained access control while enforcing a ZT security model.

The IdM, which is responsible for managing user identities and attributes, issues VCs to bind attributes to users.
These credentials are cryptographically signed by the organizations that issued them.
When requesting access to protected resources, users include the VP generated from their credential, which provides the necessary proof of their attributes.

For access control, the proposed IAM solution employs a ReBAC-based approach, where policies define the relationships that must exist between entities for access to be granted.
The PEP intercepts access requests and forwards them to the PDP for evaluation before allowing or denying access to the specified resources.
The PDP determines whether a user has the required attributes and relationships to access a specific resource by evaluating predefined access control policies. 
To achieve that, the system queries the PAP, which contains the access policies, ensuring that only users meeting the required conditions can access protected resources.
The PIP acts as the bridge between the PDP and the IdM, enabling credential verification and attribute retrieval.
Specifically, when the PDP evaluates a request, it may require additional identity information, such as whether a user’s credential is still valid or whether her VC has expired or her attributes have changed. 
The PIP interacts with the IdM 
to retrieve this information and ensure real-time validation.

\begin{figure}
    \centering
    \includegraphics[width=\linewidth]{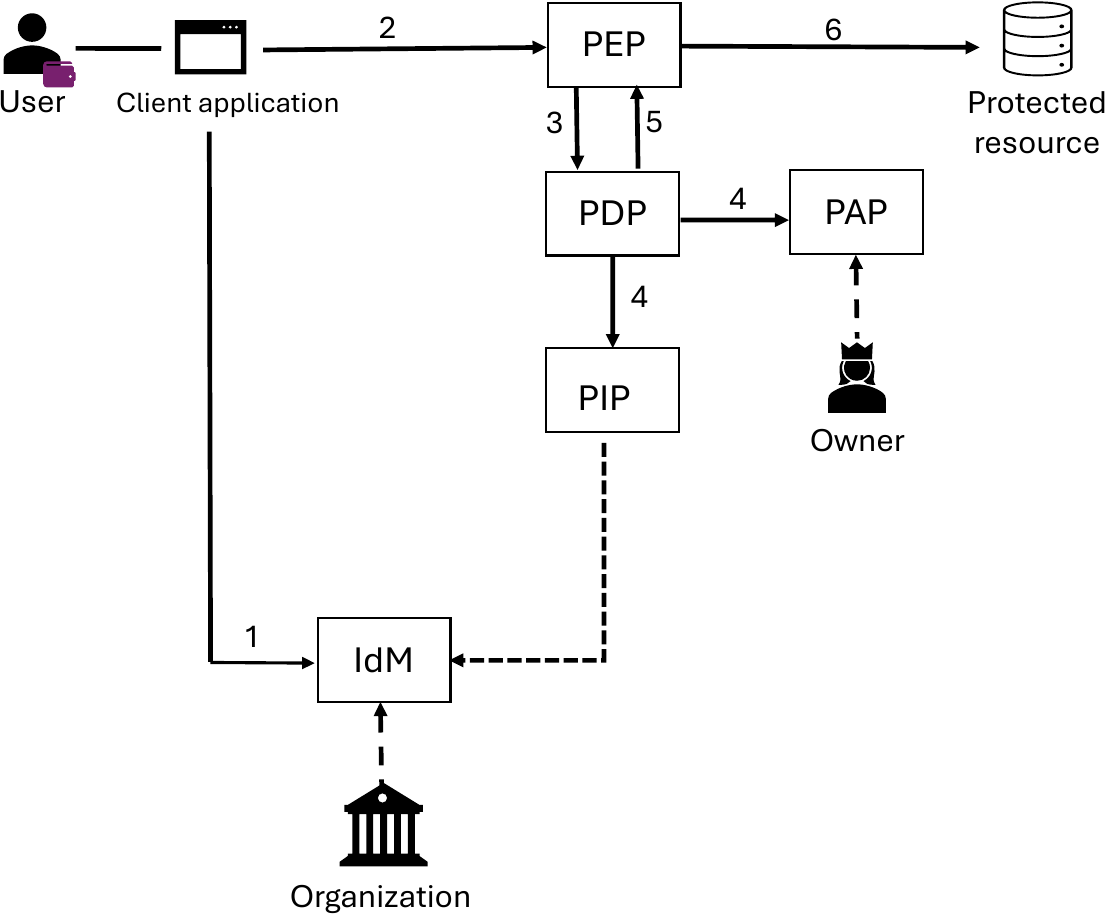} 
    \caption{Overview of the proposed IAM solution workflow.}
    \label{fig3}
\end{figure}
The workflow of the proposed IAM solution is depicted in Figure~\ref{fig3}.
Our solution assumes a setup and enrollment phase during which the identities
of the users are established. 
Both the organization and the user interact with the IdM, submitting necessary metadata and key material to receive a DID as a response. 
Organizations can then associate the DID of a user with specific attributes 

\subsubsection{VC Issuance}
After completing the enrollment process, users can request a VC  from IdM system of their organization. This VC serves as a cryptographically verifiable digital identity document that contains essential attributes about the user. These attributes typically include the user's DID, relevant roles or attributes, and any other contextual information required for access control (step 1). Additionally, the VC includes an expiration timestamp, ensuring that its validity is limited to a predefined period, and metadata that allows verification of its revocation status. This enables $3^{rd}$ parties to check whether the credential has been revoked before granting access.

Once issued, the VC is digitally signed by the organization using cryptographic mechanisms, ensuring its authenticity and integrity. The signed credential is then securely stored in the user’s digital \emph{wallet}. Since a VC is designed to have a long lifespan, it can be reused multiple times without requiring re-issuance, reducing the burden on identity providers and streamlining authentication processes.

\subsubsection{Authorization request}
When a user attempts to access a protected resource, they first generate a Verifiable Presentation (VP) derived from their VC. The VP serves as a cryptographic proof of identity and authorization, ensuring that the presented credentials are valid and controlled by the user. To establish authenticity, the VP is digitally signed using the private key associated with the user’s DID, which is included in the original VC. This signing process guarantees that only the legitimate credential holder can generate a valid VP, preventing unauthorized use by malicious actors.

To mitigate replay attacks and ensure request uniqueness, the VP incorporates a nonce, which is generated by hashing multiple dynamic elements. These include the HTTP headers of the current request, the current timestamp, and a randomly generated 128-bit number. By including these factors, the system ensures that each VP is uniquely tied to a specific session and cannot be reused by attackers. Once generated, the VP is embedded in the HTTP authentication header of the request (step~2).

\subsubsection{Authorization decision and enforcement}
Upon receiving an access request, the PEP processes the request and forwards it to the PDP for evaluation. The PDP begins by extracting the provided VP and sending it to the PIP for validation. The PIP first verifies that the VC embedded within the VP has been cryptographically signed by a trusted organization, ensuring its authenticity. Then, the PIP interacts retrieves the DID document corresponding to the user and check the revocation status of the VC. Using the obtained DID document, the PIP extracts the user’s public key and verifies the digital signature of the VP. If all validation steps succeed, the PIP returns the user’s attributes to the PDP for further processing.

With the retrieved user attributes, the PDP constructs \emph{contextual relationships}, i.e., temporary relationships that exist only during the access control decision process and are not permanently stored. These contextual relationships enable dynamic policy enforcement based on the authorization model. For example, suppose an authorization rule states that "users who have the researcher relationship with a trusted organization also inherit the reader relationship with smart plug 1". Based on the information included in a VP the system can infer a contextual relationship such as "Alice is a researcher at Technical University." The PDP then formulates an authorization query: "Does Alice have a reader relationship with smart plug 1?" Based on the defined policy, the response will be $true$, granting Alice access to the resource. An authorization decision in the returned to the PEP (step~6).

Finally, the PEP enforces the authorization decision. 
Specifically, if access is granted, the PEP forwards the request to the endpoint of the protected resource, allowing the user to interact with it. 
Otherwise, if access is denied, the PEP rejects the request, ensuring that unauthorized users cannot have access to protected resources (step~6).

\section{Implementation}
In this section, we present the implementation of our IAM framework for the cloud-to-edge continuum.
For the IdM of the proposed solution, we utilized FLUIDOS IdM~\cite{murcia2025decentralised}, which is a decentralized solution, tailored for multi-domain computing continuum frameworks and is available as open-source software.~\footnote{\url{https://github.com/fluidos-project/idm-fluidos-aries-framework-go/tree/main}}
All access control system components are provided as docker containers, as well as Helm Charts for Kubernetes-based orchestration, so as to ensure scalability and compatibility with the computing continuum.  
The source code and configurations are publicly available at an open-source repository.~\footnote{~\url{ https://github.com/excid-io/iam4cc-dev}} Note that the implementation has been validated using Minikube, which provides a controlled environment for testing Kubernetes-based integrations before transitioning to larger-scale deployments.

From a high-level perspective the system components are implemented as follows.
The PAP has been implemented as a custom .NET application that provides the necessary functionality as an HTTP REST API.  
The Helm Chart for this component includes a Kubernetes Deployment, as well as a Kubernetes Service. 
Specifically, the deployment supports horizontal scaling by adjusting the number of replicas or defaulting to one instance and uses a locally built Docker image via Minikube’s Docker daemon. 
It includes environment variables for the runtime configuration, as well as volumes and volume mounts for application settings and sensitive keys. 
The ConfigMap stores all the non-sensitive configurations, such as database connections and API endpoints, while the Secret secures sensitive information, including cryptographic keys, used at runtime. 
Relationships are stored in a private instance of OpenFGA’s open-source Helm Chart provided by the official repository.\footnote{~\url{https://artifacthub.io/packages/helm/openfga/openfga}}

Ory's Oathkeeper open-source solution~\footnote{~\url{https://www.ory.sh/docs/oathkeeper}} is employed as the PEP to intercept HTTP requests and enforce access control policies. 
The Oathkeeper Helm Chart from the official repository~\footnote{~\url{https://k8s.ory.sh/helm/oathkeeper.html}} was configured to include authenticators, authorizers, and mutators, to support the dynamic validation of VPs. 
These configurations include a JSON file that defines the access rules, mapping requests to authorization policies, while a YAML file configures authenticators, authorizers and mutators.

The PDP is also implemented as a custom .NET  application that provides an HTTP REST API that can be used to send and retrieve authorization requests. 
The PDP evaluates access requests against predefined policies stored in OpenFGA. 
Similarly to the PEP Helm Chart, the PDP Helm Chart includes a Kubernetes Deployment and a Service. 
The Kubernetes Deployment supports horizontal scaling by adjusting the number of replicas or defaulting to one instance, ensuring consistent availability. 
The Deployment includes environment variables for runtime configuration and uses a locally built Docker image via Minikube’s Docker daemon. 
A Service exposes the PDP, ensuring connectivity with all the other components of our IAM solution. 
 

\section{Evaluation}
\begin{figure}
    \centering
    \includegraphics[width=0.9\linewidth]{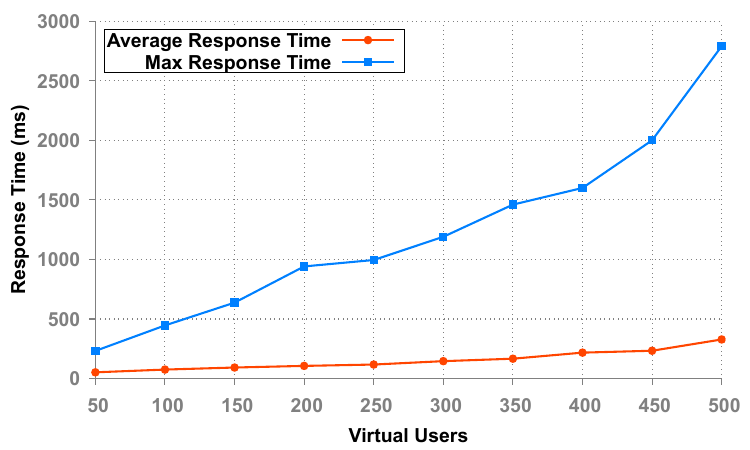} 
    \caption{Average Response Time and Max Response Time of an access control decision.}
    \label{fig4}
\end{figure}

\begin{figure}
    \centering
    \includegraphics[width=0.9\linewidth]{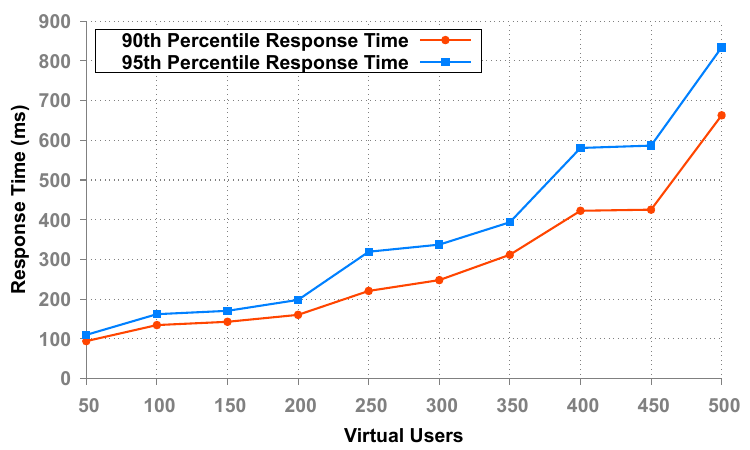} 
    \caption{ 90th and 95th Percentile Response Times of an access control decision.}
    \label{fig5}
\end{figure}

\begin{figure}
    \centering
    \includegraphics[width=0.9\linewidth]{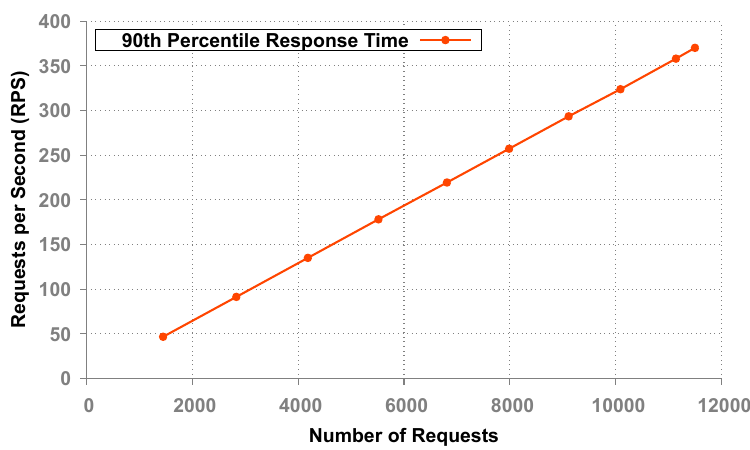} 
    \caption{ Request per second and number of requests across tests.}
    \label{fig6}
\end{figure}
To assess the performance and scalability of the proposed IAM solution we measure the time required to make and enforce an access control decision.

For the evaluation, we utilized Grafana's k6,\footnote{\url{https://k6.io/}} which is a load testing tool designed to simulate real-world request patterns and assess system responsiveness under different workloads.
The experiments were conducted in a single physical machine equipped with an Intel i7 processor, 8 CPUs, 16GB RAM, within a controlled environment for testing, utilizing Minikube. These resources were equally shared to our system's components. 



To evaluate the system components, we used \emph{Virtual Users}~(VUs) to simulate concurrent users accessing the system. 
VUs represent independent instances of simulated clients executing a workload defined in a test script within the specified duration, allowing us to replicate real-world usage patterns at varying load levels. 
Note that VUs do not inherently send requests at the exact same time; their operations are asynchronous.
For instance, 100 VUs means~100 simulated users are running the workload script simultaneously, but this does not guarantee 100 \emph{Requests per Second}~(RPS).
Each VU executes the steps in the workload script sequentially (sending a request, waiting for a response, and then proceeding to the next operation). 
Once the script completes, a VU starts over and continues until the test duration ends and thus the RPS is also dependent on the overall system load and response time.
For each component, we conduct a set of experiments that includes ten test cycles, with each of them running for 30 seconds, followed by a 30-second cool-down period to prevent unexpected results from sudden load changes.
Additionally, before running the workload test script, we define a ReBAC model and rules that establish hierarchical relationships, organizational roles, and contextual attributes, similar to the use case scenario in Section~\ref{duc}. 
This ensures that the requests made during the test require the evaluation of nested permissions, verifying the system's ability to handle complex authorization scenarios.
We use the following configuration range for VUs: low load with 50, 100 and 150 VUs to simulate light usage scenarios, moderate load with 200, 250, 300 and 350 VUs to represent typical usage levels, and high load with 400, 450 and 500 VUs to simulate peak demand.

The results for the average, maximum,~95th, and~90th percentile response times for  the PEP, including forwarding the request to the PDP, are shown in Figures~\ref{fig4},~\ref{fig5} and~\ref{fig6} respictively. 
At low load~(50-150 VUs), response times were as expected, with the~95th percentile under~171ms and the~90th under~143ms. 
The highest RPS for the low load was~135.07, with an average response time of~92.57ms and a max of~637.44ms.
At medium load~(200–350 VUs), at~350 VUs the average response time was~166.74ms, but the max response time reached~1.46s. 
The~95th and~90th percentiles stayed within~393.63ms and~311.75ms, with highest RPS recorded at~293.50.
Under high load, meaning~400–500 VUs, there was a slight decline in the system's performance. 
At the highest load, the max response time was~2.79s, however the average response time was~328.17ms, and the~95th and~90th percentiles were~834.09ms and~662.94ms respectively. 
Despite the increase in latency at 500 VUs, RPS was~370.13. 
In addition, the average and the 90th percentile response times, show that the system can maintain acceptable performance even in a constrained environment, where the system's components share their resources.

\subsection{Security properties}
Our access control solution enhances security by achieving a clear segregation of trust management responsibilities. Organizations are responsible for maintaining their own user management systems, where user attributes are stored and managed. Similarly, asset owners maintain their own authorization models, defining policies that associate these user attributes with specific access rights. This separation of concerns prevents a single entity from having excessive control over the entire trust model, reducing the risk of unauthorized access due to misconfigurations or breaches in one domain. At the same time, the use of Decentralized Identifiers (DIDs) and Verifiable Credentials (VCs) ensures that user identities and attributes are securely verifiable across different domains.

The integration of Relationship-Based Access Control (ReBAC) further strengthens security by enabling flexible and scalable policy management. By leveraging explicit relationships and inheritance, access control policies can be defined at various levels of granularity. For example, rather than assigning permissions to individual devices, an asset owner can define access rights at the level of a "group of assets" (e.g., "all devices located inside Smart Home 1"). Adding or removing a device from this group is as simple as specifying a new relation tuple, ensuring that access rights are updated automatically without requiring complex policy changes. Additionally, our solution implements continuous authentication and authorization by requiring users to generate a fresh Verifiable VP with every access request.

\section{Conclusion}

In this work, we proposed and evaluated a decentralized IAM framework for the cloud-to-edge continuum, integrating DIDs, VCs, and ReBAC to enable secure, scalable, and interoperable authentication and authorization. 
By enabling decentralized identity and trust management, and by leveraging containerized, Kubernetes-managed deployments, our approach ensures efficient, real-time access control enforcement close to protected resources, while maintaining cross-domain trust and administrative independence.  
In real-world cloud-to-edge deployments with powerful machines, the solution shows significant promise, especially when leveraging Kubernetes features such as autoscaling and replicating components to meet demand dynamically.

Future work in this direction includes the evaluation of various performance-security trade-offs. For example, PIPs can cache DID documents and revocation status of VC in order to enable faster re-authorizations, at the cost of having a time windows during which a revoked VC can be used. Similarly, we will explore other DID methods (beyond the DID method considered by the FLUIDOS IdM) that do not require communication with the registry in order to retrieve a DID document.

\begin{acks}
This work has been funded in part
by EU's Horizon 2020 Programme, through the subgrant ``Identity and Access Management for the Computing Continuum '' (IAM4CC) of project FLUIDOS, under grant agreement No. 101070473.
\end{acks}

\bibliographystyle{ACM-Reference-Format}
\bibliography{sample-base}


\end{document}